\documentclass[aps,prd,
twocolumn,nofootinbib]{revtex4-1}
\usepackage{graphicx,epsf,amssymb,amsbsy,amsfonts,amssymb,amsmath,mathtools}
\usepackage[hidelinks,breaklinks]{hyperref}

\newcommand{\eq}{\begin{equation}}
\newcommand{\eqe}{\end{equation}}
\newcommand{\g}{\gamma}

\newcommand{\G}{\Gamma}

\newcommand{\eqa}{\begin{eqnarray}}
\newcommand{\eqae}{\end{eqnarray}}

\newcommand\eea{\end{eqnarray}}
\newcommand\bea{\begin{eqnarray}}

\def\Z{\mathbb{Z}}

\def\<{\langle}
\def\>{\rangle}
\def\+{\dagger}

\def\RH{R_{\rm H}}
\def\MH{M_{\rm H}}
\def\TH{T_{\rm H}}
\def\SH{S_{\rm BH}}
\def\PBH{P_{\rm BH}}

\def\tscr{t_{\rm scr}}
\def\tev{t_{\rm evap}}

\def\LP{\ell_{\rm P}}
\def\MP{M_{\rm P}}
\def\TP{t_{\rm P}}

\def\WH{\widehat}

\def\I2{\mathbb{I}_2}

\begin{document}

\title{Black hole evaporation and semiclassicality at large $D$}

\author{Frederik Holdt-S{\o}rensen}
\email{ndx115@alumni.ku.dk}

\author{David A. McGady}
\email{mcgady@nbi.ku.dk}

\author{Nico Wintergerst}
\email{nico.wintergerst@nbi.ku.dk}

\affiliation{Niels Bohr Institute and Niels Bohr International Academy \\
17 Blegdamsvej K\o benhavn 2100, Denmark}

\begin{abstract}
Black holes of sufficiently large initial radius are expected to be well described by a semiclassical analysis at least until half of their initial mass has evaporated away. For a small number of spacetime dimensions, this holds as long as the black hole is parametrically larger than the Planck length. In that case, curvatures are small and backreaction onto geometry is expected to be well described by a time-dependent classical metric. We point out that at large $D$, small curvature is insufficient to guarantee a valid semiclassical description of black holes. Instead, the strongest bounds come from demanding that the rate of change of the geometry is small and that black holes scramble information faster than they evaporate. This is a consequence of the enormous power of Hawking radiation in $D$-dimensions due to the large available phase space and the resulting minuscule evaporation times. Asymptotically, only black holes with entropies $S \geq D^{D+3} \log D$ are semiclassical. We comment on implications for realistic quantum gravity models in $D \leq 26$ as well as relations to bounds on theories with a large number of gravitationally interacting light species.
\end{abstract}

\maketitle

Generic (non-extremal) black holes famously have both finite entropies and temperatures, which together lead to Hawking radiance/luminosity and, eventually, complete evaporation. Absent a full treatment within quantum gravity, one studies evaporation within the semiclassical approximation, where the length-scale given by Newton's constant, $\LP:= G_N^{1/(D-2)} = \MP^{-1} = \TP$, vanishes compared to the length-scales of the geometry. Geometric backreaction via quantum mechanical fluctuations then can be safely ignored, which lets the radiation to be cleanly computed. A posteriori, one then assumes the flux calculated in the semiclassical approximation to be accurate even for a finite mass black hole.

The validity of this scheme has been subject of active debate for several decades. It necessarily breaks down at the very late stages of black hole evaporation, when curvatures become large compared to the Planck scale, requiring a full treatment in quantum gravity.  On the other hand, resolution of the black hole information paradox---usually phrased as the tendency of semiclassical black holes to turn pure quantum states into mixed ones~\cite{Hawking:1976ra}, a phenomenon that would be in clear tension with basic postulates of quantum mechanics---suggests a breakdown at much earlier stages. There is compelling evidence that this should happen at the so-called Page time~\cite{Page:1993df} by which an initially classical black hole has lost roughly half its initial area via evaporation. 

In this note, we will not explore further the issue of information loss, even though we hope that our findings might provide nontrivial insights. Instead, we focus on the much more innocent question about the properties that allow for a semiclassical treatment of the \emph{early} stages of black hole evaporation. In four spacetime dimensions, this is well understood. Whenever the Schwarzschild radius is large compared to the Planck scale, a black hole is semiclassical at least up to its Page time~\cite{Page2013}. We show that when the spacetime dimension $D$ is sufficiently large, this ceases to be true for a class of ``large'' black holes.

Recent studies of the limit of a large number of space time dimensions, $D$, have led to better understandings of aspects of classical general relativity, especially for black holes~\cite{EmparanSuzukiTanabe2013, BhattacharyyaDeMinwallaMohanSaha2016}. Yet, semiclassical (and fully quantum) features of black holes~\cite{Hod2011, Hod2011a, Hod2011b, Hod:2018grr} within these new formulations of this large-$D$ limit seem to be relatively unexplored. This note is yet another partial step in this direction.

The main thrust of our work stems from the evaporation and scrambling timescales of Schwarzschild black holes in $D$-dimensions. The scrambling time is the time that a black hole needs to process and obscure infalling information \cite{HaydenPreskill2007}. It has been conjectured to saturate various bounds, bounds which are necessary for an evaporating black hole not to violate basic properties of quantum mechanics~\cite{HaydenPreskill2007,Sekino:2008he}. At large $D$, black holes coupled to $N_D$ massless modes, with Bekenstein-Hawking entropy $\SH$ and Hawking temperature $\TH$ evaporate and scramble on the timescales:
\begin{align}
\!
\!
\!
\!
\frac\tev\TP &\sim \left( \frac{4\pi}{D}\right)^{D+1/2} 
\frac{\SH}{N_D} \times \SH^{\tfrac{1}{D-2}} 
~,
\label{eq1.1}
\\
\!
\!
\!
\!
\frac\tscr\TP &= \frac{\MP}{2\pi \TH} \log \SH \sim 
\SH^{\tfrac{1}{D-2}} \times \frac{\log \SH}{D^{1/2}}~,
\label{eq1.2}
\end{align}
where we have assumed the usual expression \cite{HaydenPreskill2007,Sekino:2008he} for the scrambling time to hold in general $D$, a premise that we will discuss more towards the end of the note.
We have parametrized the timescales in terms of the entropy $\SH$ as it is the only dimensionless quantity in pure gravity in asymptotically flat spacetimes. As such, it allows us to capture the entire $D$-scaling without having to consider the individual $D$-dependence of dimensionful quantities, such as Newton's constant or the Schwarzschild radius.

Even from a very superficial view on black hole information processing, it seems clear that a black hole cannot evaporate  faster than it scrambles information. Any black hole that appears to do so cannot possibly evaporate thermally and thus be described by semiclassical physics. Yet, for fixed $\SH$, the ratio of the scrambling and evaporation times indeed increases as $\tscr/\tev \sim D^{D} N_D \log \SH/\SH$. Thus, any \emph{fixed entropy} black hole can be described by semiclassical physics only up to some critical dimension $D_{\rm crit}(\SH)$. Properly semiclassical black holes have bounded entropy:
\begin{equation}
\SH \gtrsim D^{D+3}\log D 
\,.
\end{equation}
This implies semiclassical black holes have $\tfrac\RH\LP \gtrsim D^{3/2}$. 
We stress that we work with manifestly dimensionless ratios and thus may safely ignore how $\LP = \TP = 1/\MP$ scale with the spacetime dimension, $D$.

\section{Schwarzschild in $D$-dimensions}\label{secPBH}

In this section, we first give general properties of $D$-dimensional Schwarzschild black holes. We then find their semiclassical evaporation and scrambling times.

\subsection{The metric, entropy, temperature}

The metric for the $D$-dimensional Schwarzschild black hole with ADM mass $\MH$ is well known to be,
\begin{align}
\!\!
&ds^2 = -h_D(r) dt^2 + \frac{dr^2}{h_D(r)} + r^2 d\Omega_{D-2}~,
\label{eq3.1} \\
\!\!
&h_D(r) = 1 - \left( \frac{\RH}{r}\right)^{D-3}\!\!\!\!\!\!,~\left(\frac{\RH}{\LP}\right)^{D-3} \!\!\!\!\!\!= \frac\MH\MP \frac{16 \pi/\Omega_{D-2}}{(D-2)}, \!\!\!
\label{eq3.2}
\end{align}
where $\Omega_{D-2} \equiv 2\pi^{\tfrac{D-1}{2}}/\G(\tfrac{D-1}{2})$ is the area of the $(D-2)$-dimensional unit sphere.
The periodicity properties of $g_{00}(r) = -h_D(r)$ in Euclidean signature directly give the Hawking temperature:
\begin{align}
t_E \sim t_E + \beta_{\rm BH}~~,~~\TH = \frac{1}{\beta_{\rm BH}} = \frac{(D-3)}{4\pi \RH}~.
\label{eq3.3}
\end{align}
This factor of $D$ is responsible for the high luminosities that are the focus of this note. It is straightforward to see that the semiclassical Bekenstein-Hawking entropy is:
\begin{align}
\!\!\!\!
\SH := \frac{A_{\rm BH}}{4 G_N} 
= \frac{\Omega_{D-2}}{4}\bigg(\frac{\RH}{\LP} \bigg)^{D-2} 
\!\!\!\!
= \frac{4\pi}{D-2} \RH\MH~. \!\!
\label{eq3.4}
\end{align}
The black hole's area is $A_{\rm BH}$; again $G_N := \LP^{D-2}$. We frame our main discussion in terms of the entropy, $\SH$. 

\subsection{Evaporation times and scrambling times }

We now combine properties of $D$-dimensional black holes with those of blackbodies in $D$-dimensions. Straightforward computations, for example in~\cite{EmparanHorowitzMyers2000}, give the following luminosity of a spherical blackbody with radius $R$ and temperature $T$ in $D$-dimensions:
\begin{align}
P_D = ~(N_D  D)~\frac{(TR)^D}{R^2}  ~ \frac{D-1}{D}\frac{\zeta(D)}{\pi}~ .
\label{eq2.3}
\end{align}
Apart from dimensionless factors of $D$, this expression is simple to understand. Since the power of blackbodies is proportional to their area, one has $P_D \sim R^{D-2}$. Dimensional analysis then fixes the scaling with $T$. Finally, summing over all decay channels gives a factor of $N_D$. Note that $\tfrac{D-1}{D}\zeta(D)$ rapidly goes to 1 at large $D$. 

We now use the properties of black holes as approximate blackbody radiators in $D$-dimensions. The $D$-dimensional Bose-Einstein distribution for temperature $T = \TH$ peaks near: 
\begin{align}
E \sim E^\star := (D-1) \TH = (D-1)\frac{D-3}{4 \pi \RH} \simeq \frac{D^2}{4 \pi \RH}~. 
\label{eq3.5}
\end{align}
Thus, when $D \gg 1$, black hole radiance is dominated by wavelengths small compared to $\RH$. This excises greybody physics---which we parametrize and denote by the factor $\g_D(\RH)$---from contributing to black hole radiance for even moderate values of $D$ (e.g. $D\gtrsim 8$) \cite{Hod2011}.  

There is a further, slight, modification of the absorption/emission area of the black hole. Rather than being a function of $\RH$, it is parametrized by the maximum critical impact parameter, $b_{\rm C}$, below which null rays are captured by a black hole \cite{EmparanHorowitzMyers2000}: $\tfrac{b_{\rm C}}{\RH} = (\tfrac{D-1}{2})^{1/(D-3)} (\tfrac{D-1}{D-3})^{1/2}$. 

Combining these yields the black hole luminosity,
\begin{align}
\!
\!
\!
\!
\PBH = -\frac{d \MH}{dt} = \frac{8 \pi}{e^2} \frac{N_D}{\RH^2} \bigg( \frac{D}{4\pi} \bigg)^{D+2} \!\!\!\!\!\! \times K_D \times \gamma_D(\RH)~,
\!
\!
\!
\label{eq3.6}
\end{align}
where both $K_D$, defined in Appendix~\ref{secApp} in Eq.~\eqref{eqA1}, and $\gamma_D(\RH)$ approach unity as $D$ grows. Note that the most important factor in $\PBH$, $(D/4\pi)^D$, fundamentally comes from the fact that $\RH \TH = \tfrac{D-3}{4\pi}$.

Thus, a black hole with entropy $\SH$ evaporates after
\begin{align}
\!
\!
\!
\!
\!
\frac{\tev}{\TP} &= \frac{e^2}{8\pi} \sqrt{\frac{2}{e}} \left(\frac{D}{4\pi}\right)^{-\left(D+\tfrac{1}{2}\right)}   
\frac{\SH}{N_D} \times \SH^{\tfrac{1}{D-2}} \times \frac{L_D}{K_D}  ,
\label{eq3.7}
\!
\!
\!
\end{align}
where $L_D$, defined in Eq.~\eqref{eqA2}, also goes to unity for $D$ grows. This expression for $\tev$ comes from rewriting $\MH$ in terms of $\SH$. We recover Eq.~\eqref{eq1.1} as $K_D/L_D \to 1$ and $(D-3)^2/(D(D-1)) \to 1$. 

Finally, the scrambling time is~\cite{Sekino:2008he}:
\begin{align}
\frac{\tscr}{\TP} := \frac{\MP}{2\pi \TH} \log \SH
=
2 
\frac{\log \SH}{D-3} 
\times \bigg(\frac{4 \SH}{\Omega_{D-2}}\bigg)^{\tfrac{1}{D-2}}
\!\!\!\!\!\!
~. 
\label{eq3.9}
\end{align}
As is well known, in $D = 4$ a large black hole, with $\SH \approx (\RH/\LP)^2 \gg 1$, will scramble \emph{significantly} faster than it evaporates: $\tscr \ll \tev $. However, the factor of $\SH/(\tfrac{D}{4\pi})^D$ in $\tev$ makes room for large black holes in $D \gg 1$ that have both $\SH \gg 1$ and $\tscr \gg \tev$.

\section{Semiclassical physics and large $D$}

Towards the end of the evaporation process the semiclassical analysis is expected to break down for any black hole. 
For instance, we expect that once the evaporation has proceeded to a sufficiently advanced point, the dynamics of the black hole are no longer well-described by the background spacetime evolving according solely to the classical Einstein equations. Nonetheless, for sufficiently large initial entropies, the evaporation and scrambling times are very well approximated by the expressions in Eqs.~\eqref{eq3.7} and~\eqref{eq3.9}. In this section, we make this statement more precise. In particular, we identify sources of the breakdown of semiclassicality, such as large curvature at the horizon or quasistaticity of the geometry. We will see that at sufficiently large $D$, the strongest bound is obtained from demanding scrambling times to be short as compared to the black hole lifetime.

To this end, it is useful to discuss black hole families in $D$-dimensions, indexed by positive numbers $k$ and $\WH S_0$:
\begin{align}
\!\!
\SH(k,D):= \WH S_0 D^{\tfrac{Dk}{2}} \implies
\begin{cases}
\RH \sim \LP D^{\tfrac{k+1}{2}} \\
\MH \sim \MP D^{\tfrac{Dk+1}{2}}
\end{cases}\!\!.
\label{eq3.X4}
\end{align}
At large $D$, these families of black holes (i.e. these large-$D$ limits) exactly correspond to those studied by Emparan et al \cite{EmparanSuzukiTanabe2013} (and are related to the large-$D$ limits studied by Battacharya et al \cite{BhattacharyyaDeMinwallaMohanSaha2016}), where they fix $\RH := \WH R_0 D^{\ell/2}$ and study families of black holes with $\ell = 1$ and $\ell = 2$. (Note: their $D^{\ell/2}$ is our $D^{(k+1)/2}$.)

\subsection{Where should semiclassical gravity apply?}

We need clear criteria where the above semiclassical analysis applies. Clearly, we must have $\RH > \LP$ and $\SH \gg {\cal O}(1)$. However, there are further conditions:

\paragraph*{Sub-Planckian curvature:}

First, we should require that the length-scale defined by the curvature invariant $R_{\alpha \beta \mu \nu}^2:=R_{\alpha \beta \mu \nu} R^{\alpha \beta \mu \nu}|_{r = \RH}$ to be sub-Planckian:
\begin{align}
\frac{R_{\alpha \beta \mu \nu}^2}{\MP^4}\bigg|_{r = \RH} 
\!\!
\!\!
\!\!
\!\!
\!\!
\!\!
&= \frac{(D-1)(D-2)^2(D-3)}{(\RH/\LP)^4} \notag \\
&\simeq \bigg( \frac{\sqrt{D}}{\SH} \bigg)^{\tfrac4D}
\ll 1~.
\label{eq3.X0}
\end{align}
This bound serves to ensure the subdominance of higher curvature corrections to the Einstein Hilbert action under the assumption of technical naturalness.

\paragraph*{Softness of radiation:}
Second, we fix the energy of the most likely quanta $E^\star$ to be \emph{lighter} than the black hole:
\begin{align}
\frac{E^\star}{\MH} \simeq \frac{D}{\SH} &\ll 1~.
\label{eq3.X3}
\end{align}
The radiation cannot match a blackbody if $E^\star \sim \MH$.\footnote{
We may also require $E^\star< \MP$ \cite{EmparanSuzukiTanabe2013}. However, this does not seem directly connected to the integrity of semiclassicality. Amusingly, the shallow gravitational potential in $D \gg 1$ is less efficient at attenuating the energy in transplanckian Hawking quanta created near the horizon. In $D \gg 1$, transplanckian Hawking quanta may propagate to infinity, magnifying the transplanckian problem.
}

\paragraph*{Quasistatic geometry:}
Third, we would like the black hole geometry to be relatively static during evaporation:
\begin{align}
\!
\!
\!
\!
\!
\!
\left|\frac{d \RH}{dt}\right| &= \left|\frac{d \MH}{dt} \frac{d \RH}{d\MH}\right| 
= \frac{\PBH}{D-3} \frac{\RH}{\MH}
\ll 1~.
\label{eq3.X2}
\end{align}
Explicitly, as the Schwarzschild solution is \emph{static} in Einstein gravity, if there is any appreciable departure from static geometry, characterized by $|d\RH/dt| \not \ll 1$, then the system ceases to be semiclassical. This constraint is closely related to requiring $\dot{T}/T^2 \ll 1$ for an approximately thermal emitter. However, since $\dot{T}/T^2 \sim |\dot{\RH}|/D$, this gives a considerably weaker constraint at large $D$.

Similarly, we may demand the black hole's decay width, given by its inverse lifetime, to be much smaller than its mass. However, this related constraint is much weaker than the constraint $|d\RH/dt| \ll 1$.

\paragraph*{Short scrambling times:}
In the next section, we show that there are a range of black holes that satisfy all of the above conditions, even though their scrambling times~\eqref{eq3.9} are \emph{longer} than their semiclassical half-life, $\tev/2$~\eqref{eq3.7}. If true, this would imply that information would leak out of the black hole essentially unobscured. This is clearly incompatible with semiclassical Hawking radiation, and forces us to impose the new condition
\begin{equation}
\tscr < \tev~.
\end{equation}
When $\tscr > \tev$, unitary evolution of the black hole \emph{and} its radiation-field is in tension with this rapid decay. 

\subsection{Bounds on $k$}

Here, we focus which the families of black holes defined in~\eqref{eq3.X4} can be consistent with the constraints~\eqref{eq3.X0},~\eqref{eq3.X3}, and~\eqref{eq3.X2}. First, constraints~\eqref{eq3.X0} and~\eqref{eq3.X3} together imply:
\begin{align}
\left. 
\begin{matrix}
\frac{R^2_{\alpha \beta \mu \nu}}{\MP^4} \!\sim D^{2-2k} ~~~~~~\ll 1 \\
\frac{E^\star}{\MH} ~~\!~\sim D^{1-\tfrac{(D+1)k}{2}} \ll 1 
\end{matrix}
\right\}\implies k \geq 1 ~.
\label{eq4.1}
\end{align}
If $k \in \Z_{< 1}$, then both (a) median Hawking quanta have energies \emph{greater} than the rest energy of the black hole and (b) the curvature scales are sub-Planckian. Alternatively, at large $D$, any black hole with $k \geq 1$ has sub-Planckian curvature and emits Hawking quanta whose energies are $\sim D^{1-(D+1)k/2}$ smaller than its rest energy.

Due to the rapid evaporation times (which motivated this note), one of the tightest constraints comes from requiring the size/geometry of the black hole varies slowly over time, i.e. constraint~\eqref{eq3.X2}. Relating $\RH/\MH$ to $\SH$ via Eq.~\eqref{eq3.4}, and noting $\PBH$ from Eq.~\eqref{eq3.6}, we find
\begin{align}
\bigg|\frac{d \RH}{dt}\bigg| \sim N_D\bigg(\frac{D^{1-k/2}}{4\pi \WH S_0 }\bigg)^{D}
\ll 1 
& \implies k \geq 2 ~.
\label{eq4.2}
\end{align}
Note: this bound implicitly assumes that $N_D \ll (4 \pi \WH S_0)^D$ at large $D$. In pure gravity $N_D$ counts distinct graviton polarizations, and grows quadratically: $N_D = \tfrac{D(D-3)}{2} \sim D^2$. Thus, the ``$k = 2$'' black holes evolve \emph{slowly} at large $D$; they should be well described by semiclassical physics. 

As a consequence of the above considerations, semiclassical black holes at large $D $ should have entropies that grow at least as quickly as $\SH \geq D^D$. However, it is straightforward to see that
\begin{align}
\frac{\tev}{\tscr}\bigg|_{D \gg 1} = \bigg\{\bigg( \frac{4\pi}{D}\bigg)^D \frac{1}{N_D}  \bigg\} \frac{\SH}{\log \SH} \frac{e^2}{2} ~,
\label{eq4.3}
\end{align}
where, for ease, we have used the simplified scalings in Eqs.~\eqref{eq1.1} and~\eqref{eq1.2} with properly restored order-one factors. Minimally, $N_D$ is at least $D(D-3)/2 \simeq 8 \pi^2(D/4\pi)^2$. This lets us bound $\tev/\tscr$ from above:
\begin{align}
\frac{\tev}{\tscr}\bigg|_{D \gg 1} &
\leq \bigg\{ \bigg(\frac{4 \pi}{D} \bigg)^{D+2}\bigg\} \frac{\SH}{\log \SH} \frac{e^2}{(4 \pi)^2}~.
\end{align}
Black holes, thus, with $ (\tfrac{D}{4\pi})^D \leq \SH \leq (\tfrac{D}{4\pi})^{D+3} \log D$ evaporate parametrically faster than they scramble, without violating the bounds~\eqref{eq3.X0},~\eqref{eq3.X3}, and~\eqref{eq3.X2}, derived from semiclassicality.

However, we recall that for evolution to be unitary, as happens within the semiclassical approximation, we must have $\tscr \ll \tev$. Thus, we need to impose
\begin{equation}
\SH > \bigg(\frac{D}{4\pi}\bigg)^{D+3} \log D~.
\label{eq3xyz}
\end{equation}
We observe an increase of the minimal entropy of semiclassical black holes by a factor of $D^3 \log D$ as compared to the conventional bounds listed above. This is the main point of our note.

This is good news also from a different perspective. Demanding $k \geq 2$ alone allows for black holes that have $\tev < \TP$, despite their Schwarzschild radius scaling as $\RH \sim D^{3/2} \LP$. Indeed, if $\SH:= \WH S_0\,(D/4\pi)^{D+l}$ for $0 \leq l < 5/2$, then the evaporation times vanish in the strict large-$D$ limit: 
\begin{align}
\SH:= \WH S_0 \bigg(\frac{D}{4\pi}\bigg)^{D+l} \implies \frac{\tev}{\TP} \simeq \WH S_0 \bigg(\frac{4\pi}{D}\bigg)^{5/2-l} \!\!\!\!\!\!.\!\!\!
\label{eq4.5}
\end{align}
The scrambling bound implies $l \geq 3$, such that these black holes cannot be semiclassical.

Let us end this section by noting that the effect appears less dramatic when expressed in terms of the Schwarzschild radius. The above condition becomes
\begin{equation}
\frac{\RH}{\LP} > \bigg(\frac{4}{\Omega_{D-2}}\bigg(\frac{D}{4\pi}\bigg)^{D+3} \log D\bigg)^{\tfrac{1}{D-2}}~.
\end{equation}
For $D \gg 1$, this bounds $k \geq 2$, which in turn bounds the radii of semiclassical black holes by $\RH/\LP \gtrsim D^{3/2}$. An interesting consequence of this analysis is that semiclassical black hole temperatures are now bounded from above by $\TH/\MP = (D-3)\times(4\pi \RH/\LP)^{-1} < (4\pi \sqrt{D})^{-1}$.

\subsection{Short-lived black holes and string theory}

Clearly, when $D$ is ${\cal O}(1)$, the standard conditions that the curvature scale at the horizon, or the quasi-static nature of the black hole geometry, are together strong enough to dictate whether a black hole is well-described by semiclassical physics. I.e., for ${\cal O}(1)$ values of $D$, scrambling times for such black holes are necessarily \emph{much} shorter than their evaporation times. 

Here, we would like to briefly comment that the lowest value of $D$ where $\tscr \sim \tev$ for an otherwise ``semiclassical'' Schwarzschild black hole, i.e. one whose curvature scales are sub-Planckian and whose geometry varies slowly in time (due to Hawking radiation), is significantly larger than $D = 26$. In other words, the new condition for semiclassicality in this note only applies \emph{above} the upper critical dimension for consistent string theories.

\section{Connection to previous work}

Previous work has also noted that black hole information, entropy, and evaporation has the chance to exhibit qualitatively new features at large $D$. Though recent work on the large-$D$ limit of general relativity has focused on classical features of the theory, they are explicitly aware of the high luminosity (and short time-scales) that would be associated with semiclassical, and fully quantum, gravity in large $D$.

Moreover, semiclassical aspects of black hole physics at large $D$ have been a direct focus of~\cite{Hod2011,Hod2011a,Hod2011b,Hod:2018grr}. In particular, the Hawking luminosity was explicitly found in~\cite{Hod2011}, and Refs.~\cite{Hod2011a,Hod2011b,Hod:2018grr} have discussed the relative size of black hole entropy as compared to the entropies of unbound systems of weakly gravitating matter at large $D$. 

In particular, \cite{Hod2011b} prominently discusses possible tension between basic principles of black hole entropy and properties of weakly gravitating systems when the entropy of a black hole is $\SH(D) \lesssim (D/4\pi)^{D+\ell}$ if $\ell$ lies in the interval $\ell \in (\tfrac12,1)$. In our present context, it is very amusing to note that these black holes fall squarely within the family of black holes for which the scrambling bound is the strongest. In fact, our inequality~\eqref{eq3xyz} resolves possible tension by demonstrating that a semiclassical analysis of such black holes should not be trusted.

The results of this work are similar in spirit to results derived in the context of a large number $N$ of gravitationally interacting species \cite{Dvali2010}. This does not come as a surprise. Coupling a large number of additional light degrees of freedom to gravity opens up the phase space in a similar fashion as going to large $D$. Consequently, evaporation times and scrambling times can become comparable for sufficiently large $N$, which again sets a bound on the validity of the semiclassical approximation \cite{DvaliGomez2008}. Despite the obvious similarities, there are a few key differences. The presence of large $N$ species implies a \emph{finite} renormalization of Newton's constant, extractable for example from correlators of the form $\langle T T \rangle$, with the stress tensor $T$ \cite{DvaliRedi2008}. In that case, the breakdown of semiclassicality happens at a scale which can be shown to directly correspond to the strong coupling scale~\cite{Dvali2010}. In the present context, however, since the additional degrees of freedom correspond to momentum modes of the $D$-dimensional fields,\footnote{The effect of the ${\cal O}(D^2)$ graviton polarizations is subdominant in the present context, but has a sizable impact when considering compactifications of all but $4$ of the large $D$-dimensions~\cite{Strominger1981,Bjerrum-Bohr2003}. In that case, they literally act like additional species.}  their finite contribution to the $TT$--correlator is in fact strongly suppressed \cite{Strominger1981} and the agreement of scales seems to disappear. However, naive scaling arguments suggest that for an increased number of external legs, finite contributions may indeed increase with $D$. It would be interesting to see if this could reunite the scales.

\section{Conclusions and future work}

Analyzing the first quantum corrections to black holes in a large number of dimensions, $D \gg 1$, reveals rather curious properties. The presence of this large dimensionless number has the capacity to significantly alter the naive intuition, gleaned from black holes in four and five dimensions, for the timescales in unitary evaporation. 

Indeed, in $D \gg1$, black holes temperatures grow linearly with $D$. Explicitly, $\TH \RH \sim D$, which comes from the gradient of $g_{00}(r) \sim 1 - (\tfrac{\RH}{r})^D$ at the horizon. Further, the Hawking-luminosity grows \emph{factorially} with $D$. This growth is mainly due to the growth of the available phase space with $D$. Neither of these scalings are surprising. Yet, they inexorably lead to \emph{very} short evaporation timescales for even relatively massive black holes, with $\tfrac{\MH}{\MP} \lesssim (\tfrac{D}{4\pi})^D$, which may lead to new observations.

Within this note, we have taken the conservative approach of interpreting all possible sources of tension that appear within the semiclassical approximation as pointing to the latter's demise. In particular, we have identified a range of parameters in which otherwise semiclassical black holes appear to evaporate \emph{much faster} than they scramble information. Since this is incompatible with everything we understand about the microscopic dynamics of scrambling, we have taken this as a signal that in a large number of dimensions, semiclassical physics breaks down at scales significantly larger than naively expected.

Interestingly, this realization helps to clarify nontrivial puzzles that have arisen in the literature, such as the possibility of having hyperentropic matter at sufficiently large $D$ \cite{Hod2011a} that violates the Bekenstein bound \cite{Bekenstein:1980jp}.\footnote{Note that such matter does not violate the covariant entropy bound, which is a factor of $2/(D-2)$ weaker in $D \neq 4$ \cite{Bousso2001}. It would be very interesting to see whether the fact that the Bekenstein bound is potentially violated only outside the semiclassical regime can shed light on the mismatch of the bounds.} While recent arguments argue against the validity of such conclusions on different grounds~\cite{Hod:2018grr}, we think that our picture provides a complementary and conceptually very simple explanation why such apparent tension arises outside the regime of validity of semiclassical gravity.

One of the main assumptions underlying this work is the validity of expression \eqref{eq3.9} for the scrambling time in any $D$. From a certain point of view, this may seem unnatural. After all, it implies that the entire $D$-dependence of the scrambling time lies only in the $D$-dependence of temperature and entropy. One may thus wonder whether corrections to this expression, relevant only at sufficiently large $D$, could lead to a decrease of thermalization time scales such that $\tscr < \tev$  for any $D$ and $\SH > 1$. On the other hand, the expected nonlocal nature of scrambling dynamics \cite{Sekino:2008he} appears to directly imply scrambling to be insensitive to the number of spacetime dimensions. We hope to shed more light on this issue in future work.

Before ending, we briefly speculate on two possible tools to more directly understand and analyze the unitarity of black hole evaporation in these large-$D$ contexts. 

First, although holographic dualities between conformal field theories (CFTs) in $D$-dimensions and gravitational theories in $(D+1)$-dimensions are not expected to be ``nice'' (or, perhaps, to exist at all) when $D \gg 1$, there are hints that CFTs might be very simple at large $D$~\cite{Fitzpatrick:2013sya}. It might be interesting if this has something to say about small black hole evaporation in $(D+1)$-dimensions. 

Second, in~\cite{Emparan:2013xia}, black holes with $\RH/\LP \simeq D$ (i.e. $k =1$ or equivalently $\tev \simeq (\tfrac{4\pi}{D})^{D/2}$) were found to have a universal near-horizon limit dictated by a string theory. It would be very interesting to see if this stringy description can aid in a more direct understanding of the scrambling dynamics of semiclassical black holes. 

\acknowledgments 
We thank Peter Denton, Roberto Emparan, Troels Harmark, Cindy Keeler, Niels Obers and Marta Orselli for stimulating discussions and important comments on the manuscript, as well as Gia Dvali and Oriol Pujolas for useful discussions and Nima Arkani-Hamed for emphasizing the fundamental importance of $\tscr < \tev$.
The work of NW was supported by FNU grant number DFF-6108-00340.

\appendix

\section{$K_D$ and $L_D$}\label{secApp}

The function $K_D$ 
introduced in
expression
~\eqref{eq3.6}
for black hole lifetimes is
explicitly given by, 
\begin{align}
\!\!\!\!\!\!
K_D&:=2 e^2 \zeta(D) \bigg(\frac{D-1}{D^2}\bigg) \bigg( \frac{b_{\rm crit}}{\RH} \bigg)^{D-2} \bigg(\frac{D-3}{D} \bigg)^D \!\!.\!\!\!
\label{eqA1}
\end{align}
Similarly, the function $L_D$ introduced in expression~\eqref{eq3.7} is given explicitly by
\begin{align}
L_D &:= \bigg(\frac{D-3}{D-1}\frac{D-2}{D}\bigg) \sqrt{\frac{e}{2}}~\sqrt{\frac{4\pi}{D}}\bigg(\frac{4}{\Omega_{D-2}}\bigg)^{\tfrac{1}{D-2}}
\label{eqA2}\!\!\!\!\!\!.\!\!
\end{align}
Note: $L_D,K_D \to 1$ as $D \to \infty$. While we do \emph{not} give an expression for $\gamma_D(\RH)$, we recall $\g_D(\RH) \to 1$ 
as $D \to \infty$.

\bibliography{bibliography}

\end{document}